\documentclass[
twocolumn,
amsmath,amssymb,
aps,
prl]{revtex4-2}

\usepackage{xspace}
\usepackage{graphicx}
\usepackage{bm}
\usepackage[colorlinks=true,urlcolor=blue,linkcolor=blue,citecolor=blue,bookmarks=false]{hyperref}
\usepackage{amsmath}
\usepackage{amssymb}
\usepackage{bbold}
\usepackage{soul}
\usepackage{dsfont}
\usepackage{braket}
\usepackage{textcomp}

\newcommand{\customref}[2]{\hyperref[#1]{\ref*{#1}#2}}

\usepackage{xcolor}

\definecolor{Ured}{HTML}{cc0000}
\definecolor{Ublue}{HTML}{1f65cf}
\definecolor{Ugreen}{HTML}{198a11}

\newcounter{myequation}
\makeatletter
\@addtoreset{equation}{myequation}
\makeatother

\newcounter{myfigure}
\makeatletter
\@addtoreset{figure}{myfigure}
\makeatother

\newcommand{\ie}[0]{i.e.\@\xspace}

\newfont{\tensy}{cmsy10}

\newcommand{\fan}[1]{\hat{c}^{\vphantom\dagger}_{#1}}
\newcommand{\fcr}[1]{\hat{c}^{\dagger}_{#1}}
\newcommand{\fden}[1]{\hat{n}_{#1}}

\newcommand{\Q}[1]{\hat{Q}_{#1}}
\renewcommand{\P}[1]{\hat{P}_{#1}}

\newcommand{\fcohan}[1]{c_{#1}}
\newcommand{\fcohcr}[1]{\bar{c}_{#1}}
\newcommand{\fcohmeasure}{\mathcal{D}(\bar{c},c)}

\newcommand{\kF}{k_{\text{F}}}

\newcommand{\im}{\mathrm{i}}

\newcommand{\Hc}{\mathrm{H.c.}}

\begin{document}

\title{Intermediate bond-order-wave phase and nature of the order-to-order transition\\in the one-dimensional Hubbard-Holstein model}

\author{Manuel Weber}
\email{manuel.weber1@tu-dresden.de}
\affiliation{Institut f\"ur Theoretische Physik and W\"urzburg-Dresden Cluster of Excellence ct.qmat, Technische Universit\"at Dresden, 01062 Dresden, Germany}

\date{\today}

\begin{abstract}
The Hubbard-Holstein model is one of the central models that describe the competition between electron-electron and electron-phonon interactions. In one dimension and at half-filling, the interplay between an electronic spin-density wave and a phonon-driven charge-density wave is considered to stabilize an intermediate Luther-Emery liquid. Here we show that, once the extended metallic regime disappears, a narrow bond-order-wave phase emerges. We use an exact directed-loop quantum Monte Carlo method for retarded interactions to simulate system sizes of several hundred sites, necessary to observe the weak signatures of this novel phase. Our results suggest a second-order quantum phase transition between the two charge orders that only turns first-order at strong coupling. Our findings are reminiscent of the extended Hubbard model, but they are driven by competing frequency dependencies of the same dynamically-screened Hubbard interaction.
\end{abstract}

\maketitle

\textit{Introduction.}---%
Quantum materials can exhibit a variety of ordered phases like
antiferromagnetism, superconductivity, or different charge orders 
which often compete with each other as an external parameter is tuned
 \cite{Sipos:2008aa, Keimer:2015aa, Cao:2018ab, Wang:2023aa}.
 However, it is not always clear if they appear
as a result of electron-electron interactions, electron-phonon coupling,
or from the interplay of both. One way to identify possible mechanisms to stabilize a
certain type of order is by studying simple microscopic model Hamiltonians.
A central platform to explore the effects of strong correlations are
one-dimensional (1D) models, because---besides their relevance for quasi-1D materials \cite{GrunerBook, POUGET2016332}---%
they can be studied with great accuracy due to the availability of
powerful analytical and computational approaches.

The competition between different parts of the long-range Coulomb interaction
has been studied using the 1D half-filled \textit{extended Hubbard model}.
While the local Hubbard repulsion leads to a spin-density wave (SDW) with one electron per site,
the additional nearest-neighbor repulsion favors the formation of a charge-density wave (CDW) with strongly-bound electron pairs at every other site.
Surprisingly, the competition between both interactions stabilizes a narrow intermediate bond-order-wave (BOW) phase, for which
the electron density is enhanced on every other link between lattice sites
\cite{doi:10.1143/JPSJ.68.3123, PhysRevB.61.16377, PhysRevB.65.155113, 
PhysRevLett.91.089701, PhysRevLett.92.236401, PhysRevLett.88.056402,  PhysRevB.69.035103, PhysRevLett.92.246404, PhysRevLett.99.216403, PhysRevLett.96.036408}.
The three phases are illustrated as insets in Fig.~\ref{fig:phase diagram}.

Charge-ordered phases naturally occur from a coupling to phonons via the Peierls instability \cite{Peierls}: a periodic lattice distortion
is accompanied by CDW or BOW order, depending on whether the local lattice displacement couples to the local electron density, as in the Holstein model \cite{HOLSTEIN1959325, PhysRevB.27.4302},
or modulates the hopping integral, as in the Su-Schrieffer-Heeger (SSH) model \cite{PhysRevLett.42.1698, PhysRevB.27.1680}.
These orders can already be understood from a mean-field treatment of the phonons which becomes exact in the limit of zero phonon frequency.
Recently, this simple rule of thumb has been challenged by the study of SSH models, where the coupling to the off-diagonal bond density
can induce CDW order for spinless fermions \cite{PhysRevResearch.2.023013} or 2D SDW order  \cite{Beyl2020, PhysRevLett.127.247203, PhysRevB.105.085151} 
if quantum lattice fluctuations become sufficiently strong. By contrast, retardation effects might only 
destroy the CDW state in favor of a disordered phase if the phonons couple to the diagonal electron density \cite{PhysRevB.60.7950}.

\begin{figure}[t]
\includegraphics[width=0.755\linewidth]{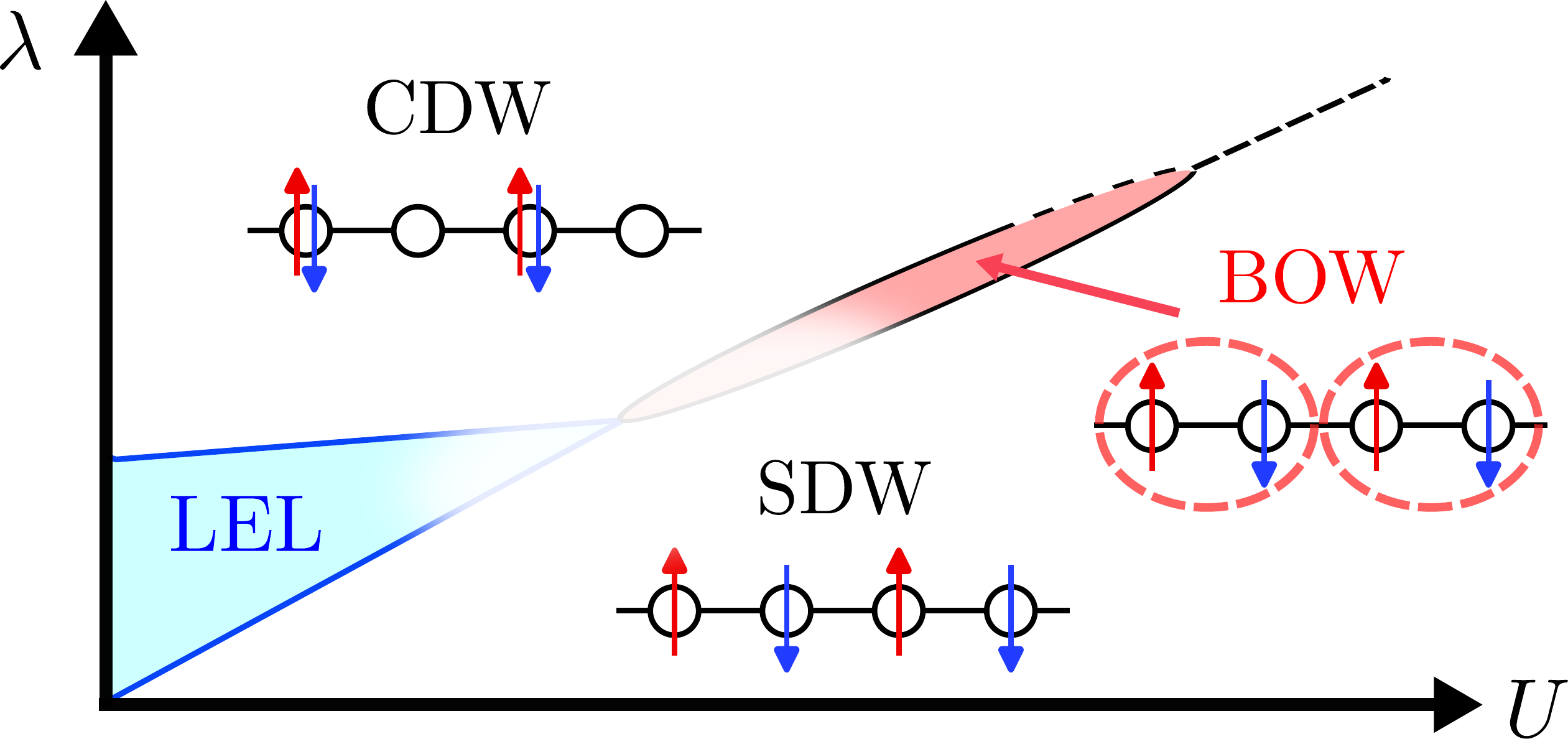}
\caption{%
Schematic phase diagram of the 1D Hubbard-Holstein model
in terms of the electron repulsion $U$ and the electron-phonon coupling $\lambda$,
as suggested by our QMC study at a finite phonon frequency of $\omega_0/t = 1$ and strong couplings.
Insets show SDW, BOW, and CDW patterns for a four-site system.
}
\label{fig:phase diagram}
\end{figure}

The competition between electronic SDW and phonon-driven CDW states is usually studied using the \textit{Hubbard-Holstein model}
\cite{PhysRevLett.69.1600, PhysRevB.52.4806, PhysRevLett.75.2570, HOTTA19971037, Fehske:2003aa, PhysRevB.69.165115, 2004EL.....66..559K, PhysRevB.75.014503, 2010EL.....9027002B, PhysRevB.81.235113, PhysRevLett.109.246404, PhysRevB.87.235133, PhysRevB.88.125126, PhysRevB.92.195102, PhysRevB.91.235150, PhysRevLett.119.197001, PhysRevB.96.205145, Costa:2020aa, PhysRevResearch.2.043258, PhysRevLett.125.167001, PhysRevB.107.235120}.
In 1D and at half-filling, it hosts an intermediate metallic regime whose nature has been debated controversially 
\cite{PhysRevB.60.7950, PhysRevB.67.081102, PhysRevLett.95.096401, PhysRevLett.95.226401, PhysRevB.75.245103, PhysRevB.76.155114, PhysRevB.75.161103, Fehske_2008, 2010JPhCS.200a2031E, PhysRevB.84.165123}
but which, by now, is considered 
\cite{PhysRevB.87.075149, PhysRevB.91.085114, PhysRevB.87.075149, PhysRevB.92.245132, PhysRevB.107.075142}
a Luther-Emery liquid (LEL) \cite{PhysRevLett.33.589}.
Its exponentially small spin gap leads to an extremely slow crossover at weak couplings
which complicates numerical verification on finite chains and the estimation of precise phase boundaries \cite{PhysRevB.92.245132}. 
The LEL is sometimes regarded as the equivalent to the intermediate phase in the extended Hubbard model \cite{PhysRevB.75.245103}, beyond which a direct SDW--CDW transition is expected to occur
\cite{FEHSKE2002562, PhysRevB.75.245103}.

In this Letter, we uncover a second intermediate phase in the 1D Hubbard-Holstein model which exhibits BOW order and occurs at strong coupling. A schematic phase diagram is shown in Fig.~\ref{fig:phase diagram}. 
We use an exact directed-loop quantum Monte Carlo (QMC) method for retarded interactions \cite{PhysRevLett.119.097401} that reaches significantly larger system sizes
than previous approaches, which is necessary to unambiguously identify this narrow phase. 
Our results suggest a second-order BOW--CDW quantum phase transition that can be interpreted as a 1D realization of a deconfined quantum critical point \cite{PhysRevB.99.165143, PhysRevB.99.205153} and only turns first-order at stronger interactions before the BOW phase disappears.
Our findings are reminiscent of the extended Hubbard model,
but in our case the BOW phase is induced by competing contributions of a frequency-dependent density-density interaction that is local in space.

\textit{Model.}---%
The 1D Hubbard-Holstein model is defined as
\begin{align}
\nonumber
\hat{H}
	=
	& -t \sum_{i\sigma} \left( \fcr{i,\sigma} \fan{i+1,\sigma} + \Hc \right)
	+\frac{U}{2} \sum_i \left( \fden{i} -1 \right)^2
\\
	 &+\sum_i \left(  \frac{1}{2m} \P{i}^2 +  \frac{k}{2}  \Q{i}^2 \right) 
	+ g \sum_i \Q{i} \left(\fden{i} - 1 \right) \, .
\label{eq:H}
\end{align}
Here $\fcr{i,\sigma}$ ($\fan{i,\sigma}$) creates (annihilates) an electron at lattice site $i\in\{1, \dots, L\}$ and with spin $\sigma\in\{ \uparrow, \downarrow\}$, whereas the phonons are written in terms of momentum and displacement operators $\P{i}$ and $\Q{i}$, respectively. The first term describes the
nearest-neighbor hopping of electrons with amplitude $t$, the second the local Hubbard repulsion with $U>0$, the third dispersionless phonons with frequency $\omega_0 = \sqrt{k/m}$,
and the fourth the coupling between the electronic density
$\fden{i}=\sum_\sigma \fcr{i,\sigma} \fan{i,\sigma}$
and the phonon displacement.
We define the dimensionless electron-phonon coupling $\lambda = g^2 / (Wk)$
using the band width of free electrons, $W=4t$.

The phonons can be integrated out exactly using the coherent-state path integral, such that the partition function
$Z = Z_\mathrm{ph}^0 \int \fcohmeasure \, e^{- \mathcal{S}[\fcohcr{}, \fcohan{}]}$
only depends on the electronic action
$\mathcal{S} = \mathcal{S}_\mathrm{el} + \mathcal{S}_\mathrm{ret}$.
The
retarded interaction
$
\mathcal{S}_\mathrm{ret}
	=
	-\frac{\lambda W}{2} \iint_0^\beta d\tau d\tau' \sum_i [n_i(\tau)-1] D(\tau-\tau') [n_i(\tau')-1] 
$
is nonlocal in imaginary time and
mediated by the free-phonon propagator $D(\tau) \propto e^{-\omega_0 \tau}$;
here
$\beta$ is the inverse temperature.
In combination with the local repulsion $U$, we obtain the frequency-dependent Hubbard interaction
$U(\omega) = U -  {\lambda W}/[{1-(\omega/\omega_0)^2}]$.
For $\omega_0 \to \infty$, the Holstein interaction maps to the attractive Hubbard model and
we define the effective Hubbard interaction $U_\infty = U - \lambda W$, but for any finite $\omega_0$
the frequency dependence becomes important. $U(\omega)$ can
also describe the dynamical screening of the Hubbard interaction \cite{PhysRevB.70.195104, PhysRevB.74.125106, PhysRevLett.104.146401, Werner_2016}.

\textit{Method.}---%
We use an exact QMC method based on a diagrammatic expansion of the partition function  $Z/Z^0_\mathrm{ph}$
 in the full fermionic action $\mathcal{S}$.
 The global directed-loop updates
allow for an efficient sampling of fermionic world-line configurations
\cite{PhysRevB.59.R14157, PhysRevE.66.046701}.
These updates readily apply to the 1D Hubbard model \cite{Sandvik1992AGO, PhysRevB.65.155113}
and have recently been extended to the retarded interaction $\mathcal{S}_\mathrm{ret}$ \cite{PhysRevLett.119.097401}
to overcome the autocorrelation problem of previous approaches that only used local updates to sample the phonons directly \cite{PhysRevB.75.245103}.
Its efficient $\mathcal{O}(L \beta \log \beta)$ scaling \cite{PhysRevLett.119.097401} allows us to reach significantly larger systems than previous studies.
For our analysis, we calculate the real-space correlation functions, structure factors, and susceptibilities \cite{PhysRevB.56.14510},
\begin{gather}
C_O(r) = \sum_i \langle \bar{O}_i \bar{O}_{i+r}  \rangle / L \, , \quad
S_O(q) = \sum_{r} e^{\im q r} C_O(r) \, ,
\\
\chi_O(q) = \frac{1}{L} \sum_{ij} e^{\im q(i-j)} \int_0^\beta d\tau \, \langle \bar{O}_i(\tau) \bar{O}_j(0) \rangle \, .
\end{gather}
We define $\bar{O}_i= \hat{O}_i - \langle \hat{O}_i \rangle$ and choose $\hat{O}_i \in \{\fden{i}, \hat{S}^x_i, \hat{B}_i\}$
for the charge, spin, and bond sectors labeled by $\rho / \sigma / b$, respectively. We also define the spin and bond operators, $\hat{S}^x_i = (\fcr{i,\uparrow} \fan{i,\downarrow} + \Hc)$
and $\hat{B}_i = \sum_\sigma (\fcr{i,\sigma} \fan{i+1,\sigma} + \Hc)$.

\textit{Results.}---%
We study the Hubbard-Holstein model at a phonon frequency of $\omega_0 / t =1$,
set $t$ as the unit of energy, scale the inverse temperature as $\beta t = 4L$ to access the ground state, and use periodic boundary conditions.
Unless noted otherwise, the electron repulsion is $U/t = 6.0$.

The possible phases of our model can be distinguished by their charge and spin gaps.
The SDW (LEL) phase has a nonzero charge (spin) gap but no spin (charge) gap, whereas
both gaps are nonzero in the CDW and BOW phases. These excitation gaps are directly related
to the Luttinger parameters $K_{\rho/\sigma}$, which determine the long-distance decay of correlation functions
and are zero if the corresponding channel is gapped.
In Luttinger-liquid theory, a finite-size estimate
can be obtained from the long-wavelength limit of the charge/spin structure factor,
\begin{align}
\label{eq:K}
K_{\rho/\sigma}(L) = \pi \, S_{\rho/\sigma}(q_1) / q_1
\, ,
\qquad
q_1 = 2\pi/L \, ,
\end{align}
from which we get $K_{\rho/\sigma} = \lim_{L\to\infty} K_{\rho/\sigma}(L)$
\cite{PhysRevB.59.4665}.

\begin{figure}[t]
\includegraphics[width=\linewidth]{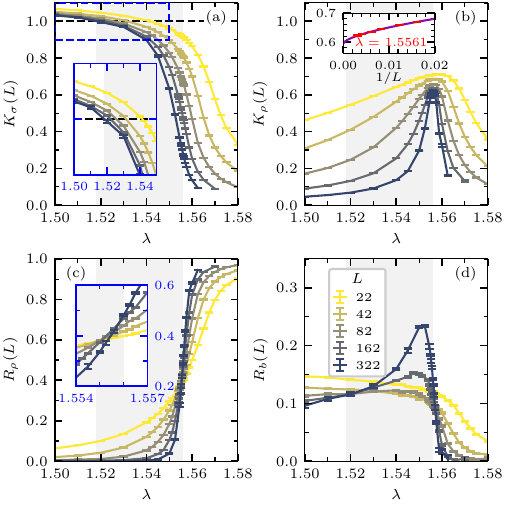}
\caption{%
Finite-size dependence of the (a) spin and (b) charge Luttinger parameters $K_{\sigma / \rho}(L)$
as well as of the (c) charge and (d) bond correlation ratios $R_{\rho / b}(L)$
across the SDW--BOW--CDW transitions. The shaded areas indicate the intermediate phase.
The insets in (a),(c) zoom into the regime 
near $\lambda_{c_1}$,$\lambda_{c_2}$,
whereas the inset in (b) extrapolates $K_\rho(L) \to 0.59(1)$ using a power-law fit \cite{PhysRevLett.92.236401}.
}
\label{fig:FSU6}
\end{figure}

Figures \ref{fig:FSU6}(a) and \ref{fig:FSU6}(b) show $K_{\rho/\sigma}(L)$ 
for different system sizes and as a function of the electron-phonon coupling $\lambda$.
For $\lambda \lesssim U/W$, our system is in the SDW phase, for which spin-rotational symmetry requires $K_\sigma = 1$.
However, $K_\sigma(L)$ is known to converge only slowly from above
due to logarithmic corrections \cite{PhysRevLett.92.236401, PhysRevLett.73.332}. The opening of a spin gap can be estimated
from the coupling $\lambda_{c_1}(L)$ at which $K_\sigma(L)$ drops below one \cite{PhysRevB.65.155113}, as depicted in Fig.~\ref{fig:FSU6}(a),
and we extrapolate
$\lambda_{c_1} = 1.518(3)$.
In contrast, 
the charge sector is gapped for all $\lambda \neq \lambda_{c_2} = 1.5560(5)$
where $K_\rho(L)$ exhibits a peak that becomes narrower with increasing $L$ 
but whose height converges to a constant. 
In the inset of Fig.~\ref{fig:FSU6}(b)
we extrapolate $K_\rho(L) \to 0.59(1)$  
at fixed $\lambda$ using a power-law fit \cite{PhysRevLett.92.236401}
but note that an exact fitting function is not known.
At this isolated point, our system has a finite spin gap but zero charge gap and therefore is in an LEL state. For $\lambda > \lambda_{c_2}$, our system is CDW ordered, but we find an additional extended regime $\lambda_{c_1} < \lambda < \lambda_{c_2}$ where spin and charge gaps are nonzero.

To identify the two spin- and charge-gapped phases, we calculate the correlation ratio
at the ordering vector,
\begin{align}
R_{\rho / b}(L) = 1- \frac{S_{\rho / b}(q=\pi + q_1)}{S_{\rho / b}(q=\pi)} \, ,
\end{align}
for the charge and bond channels. $R_{\rho / b}$ is defined in such a way that it scales to one in the ordered phase, to zero in the disordered phase, and becomes scale invariant at a critical point. 
The onset of CDW order is clearly visible from $R_\rho(L) \to 1$ in 
Fig.~\ref{fig:FSU6}(c). We find a clear crossing of $R_\rho(L)$ for data pairs $(L,2L-2)$
which drifts only slowly with increasing $L$,
hence signaling a second-order phase transition at $\lambda = \lambda_{c_2}$ consistent with our previous estimate.
The BOW phase can be inferred from the bond correlation ratio $R_b$ in Fig.~\ref{fig:FSU6}(d).
$R_b(L) \to 0$ in the CDW phase, whereas $R_b(L)$ starts to increase for $\lambda \lesssim \lambda_{c_2}$ once $L\gtrsim 82$.
Then, $R_b(L)$ also exhibits a crossing at $\lambda \approx \lambda_{c_2}$.
The absolute value of $R_b$ only scales
slowly towards one because the BOW order parameter is much smaller than the one for the CDW phase
which can already be detected on small systems.
The enhancement of $R_b$ is weaker near $\lambda_{c_1}$ because we expect a Berezinskii-Kosterlitz-Thouless (BKT) transition at the SDW--BOW border
with an exponentially small spin gap in the BOW phase \cite{PhysRevLett.92.236401}.
For $\lambda < \lambda_{c_1}$, $R_b(L)$ scales to zero only slowly, because the SDW phase has  critical spin and bond correlations.

\begin{figure}[t]
\includegraphics[width=\linewidth]{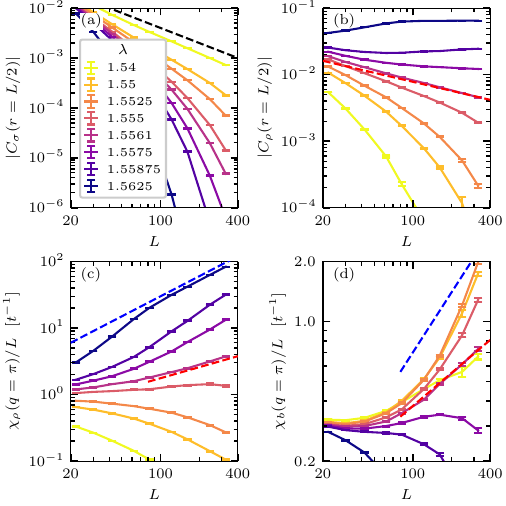}
\caption{%
(a),(b) Spin/charge correlations $C_{\sigma / \rho} (r=L/2)$ and
(c),(d) charge/bond susceptibilities $\chi_{\rho / b}(q=\pi)/L$ as a function of $L$. 
The dashed line in (a) corresponds to $L^{-1}$, the blue lines in (c),(d) to $L$, and
the red lines to $C_\rho(r=L/2) \propto L^{-K_\rho}$ and $\chi_{\rho/b}(q=\pi)/L \propto L^{1-K_\rho}$ scaling
with $K_\rho\approx 0.45$.
}
\label{fig:realspace_sus_U6}
\end{figure}

The long-distance decay in the spin and charge sectors is displayed in Figs.~\ref{fig:realspace_sus_U6}(a) and \ref{fig:realspace_sus_U6}(b)
via $\left| C_{\sigma / \rho}(r=L/2) \right|$ as a function of system size.
For $\lambda > \lambda_{c_1}$, the spin gap leads to an exponential decay of $\left| C_\sigma(r=L/2) \right|$,
which slowly crosses over towards the characteristic $1/r$ behavior of the SDW phase once the exponentially small spin gap cannot be resolved anymore on a finite system.
By contrast, $\left| C_\rho(r=L/2) \right|$ goes to a constant in the CDW phase, but decays exponentially for $\lambda < \lambda_{c_2}$.
The charge-ordered phases can also be identified from the charge/bond susceptibility
$\chi_{\rho / b}(q=\pi)/L$ depicted in Figs.~\ref{fig:realspace_sus_U6}(c) and \ref{fig:realspace_sus_U6}(d). In the CDW/BOW phase, we expect $\chi_{\rho/b}(q=\pi)/L \propto L$ and $\chi_{b/\rho}(q=\pi)/L \to 0$. Deep in the CDW phase, $\chi_{\rho}(q=\pi)/L$ first diverges faster than $L$ before it reaches its asymptotic linear $L$ dependence. In the BOW phase, we observe the same overshooting in $\chi_b(q=\pi)/L$, which is a strong signature of BOW order, but systems are too small to observe convergence to the asymptotic behavior.
At the BOW--CDW phase boundary, the existence of an LEL fixed point requires
$C_\rho(r) \propto (-1)^r r^{-K_\rho}$ and, using conformal invariance,
$\chi_{\rho / b}(q=\pi)/L \propto L^{1-K_\rho}$. For the largest systems, 
we observe a common power-law exponent at $\lambda = 1.5561$ (we use $K_\rho \approx 0.45$ as a guide to the eye), but size effects
are still strong in the critical regime.

\begin{figure}[t]
\includegraphics[width=\linewidth]{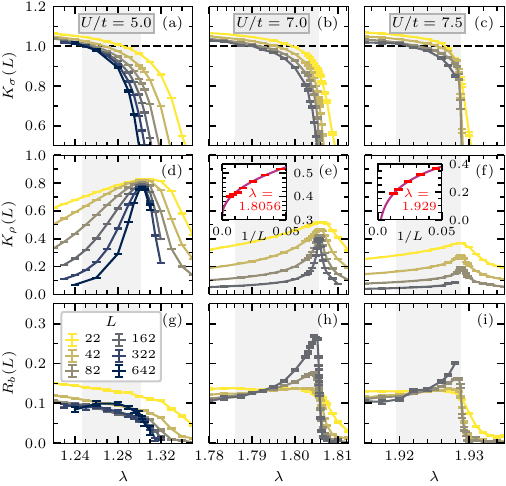}
\caption{%
Finite-size dependence of the (a)--(c) spin and (d)--(f) charge Luttinger parameters $K_{\sigma / \rho}(L)$ as well as of the (g)--(i) bond correlation ratio $R_b(L)$ for different Hubbard repulsions $U/t \in \{5.0, 7.0, 7.5 \}$ listed on top of each column. The insets in (e) and (f) extrapolate the maximum of $K_\rho(L)$; we include power-law fits as a guide to the eye.
}
\label{fig:differentU}
\end{figure}

Finally, the effects of the electron repulsion $U$
are studied in Fig.~\ref{fig:differentU}. 
As before,
we can identify an intermediate regime $\lambda_{c_1} < \lambda < \lambda_{c_2}$  from $K_{\sigma/\rho}(L)$ shown in Figs.~\ref{fig:differentU}(a)--(f) whose extent shrinks if we increase $U/t$ from $5.0$ to $7.5$ and which eventually disappears for $U/t \gtrsim 8.0$ \cite{SM_HubHolBOW}.
The BOW response of the intermediate phase is strongest at $U/t=7.0$ in Fig.~\ref{fig:differentU}(h), where $R_b(L)$ increases for all $L \gtrsim 22$,
whereas an enhancement of $R_b(L)$ is already hard to detect at $U/t=5.0$ in Fig.~\ref{fig:differentU}(g).
At $U/t=7.5$ in Fig.~\ref{fig:differentU}(f), we find $K_\rho(L) \to 0$ which is a direct signature of a first-order BOW--CDW transition.

\textit{Discussion.}---%
Our findings at strong interactions are in close analogy to the extended Hubbard model.
The latter hosts an intermediate BOW phase which exhibits a BKT transition to the
SDW phase and a second-order transition to the CDW state that turns first-order at strong coupling. The BOW--CDW transition is understood from a Gaussian theory for the LEL fixed point \cite{PhysRevLett.92.236401}: 
The LEL becomes unstable towards $4\kF$ Umklapp scattering once $K_\rho < 1$. Then, the sign of the corresponding coupling parameter determines if the system flows to the CDW or BOW state.
Exactly at the BOW--CDW transition, this coupling is tuned through zero such that the LEL phase remains stable for $1/4 < K_\rho < 1$ until the $8\kF$ Umklapp term becomes relevant \cite{PhysRevLett.92.236401}. Because the LEL has gapless charge excitations, it describes a second-order transition.

Overall, our results for the 1D Hubbard-Holstein model are consistent with this picture:
For both models, $K_\rho(L)$ shows a peak at the BOW--CDW phase boundary
that becomes more narrow with increasing $L$ and converges to a finite $K_\rho < 1$ \cite{PhysRevB.65.155113, PhysRevLett.92.236401}.
By raising the electron repulsion in Fig.~\ref{fig:differentU} up to $U/t = 7.0$, $K_\rho$ decreases until it drops to zero for $U/t=7.5$ in Fig.~\ref{fig:differentU}(f). 
At the BOW--CDW transition, charge and bond correlations show similar power-law exponents in Fig.~\ref{fig:realspace_sus_U6} that deviate from $K_\rho(L\to\infty)$ in Fig.~\ref{fig:FSU6}(b), though.
This discrepancy is likely a result of crossover effects which are strong when the BOW phase emerges
and the spin gap cannot be fully resolved yet \cite{PhysRevB.65.155113};
in particular,
correlation functions are sensitive to finite-size effects \cite{SM_HubHolBOW} 
and small detunings from $\lambda_{c_2}$, 
but $K_\rho(L)$ also shows a drift that is difficult to extrapolate.
It has also been debated if the
coupling to phonons can lead to further complications in the charge sector
\cite{PhysRevB.75.245103, PhysRevB.84.165123, PhysRevB.92.245132}.

The second-order BOW--CDW phase transition can be
interpreted in terms of 1D deconfined criticality \cite{PhysRevB.99.165143, PhysRevB.99.205153}.
In 1D electron-phonon models, 
such a transition can occur from 
competing site and bond phonons \cite{PhysRevLett.117.206404} or from varying the retardation range when only coupling to bond phonons \cite{PhysRevResearch.2.023013}. 
Our work shows that such a transition can even arise from the frequency dependence of a purely local density-density interaction as a direct consequence of  quantum lattice fluctuations. It is absent for 
$\omega_0 \to \infty$, where the resulting Hubbard model exhibits an SDW--LEL transition at $U_\infty = 0$,
as well as for
$\omega_0 \to 0$,
where the LEL fixed point of the phase boundary is always unstable towards static dimerization \cite{10.1093/acprof:oso/9780198525004.001.0001, PhysRevB.110.125130}.

For $U/t < 5.0$, even $L=642$ sites are not big enough to find a direct signature of BOW order in $R_b(L)$. 
Nonetheless, our procedure to estimate the regime $\lambda_{c_1} < \lambda < \lambda_{c_2}$ from $K_{\rho/\sigma}(L)$
suggests that our intermediate phase extends to much weaker $U$ \cite{SM_HubHolBOW}.
It is most natural to assume that our LEL phase boundary emerges from the tip of the extended LEL regime, as illustrated in Fig.~\ref{fig:phase diagram},
once Umklapp scattering becomes relevant at a given $K_\rho$. 
However, it is difficult to obtain a direct signature of the LEL phase at weak coupling,
where numerical simulations on finite chains can only detect crossover behavior in correlation functions \cite{PhysRevB.92.245132}.
The same is true for the attractive Hubbard model at $U_\infty \gtrsim 0$, 
such that the frequency dependence of $U(\omega)$ needs to drive the crossover at $\omega_0 < \infty$.
Large $\lambda$ are expected to speed up this process,
but we still see signatures of a crossover in our results. 
At $\omega_0/t =1$, the existence of an extended LEL phase is supported by a one-loop 
renormalization-group study \cite{PhysRevB.91.085114} which,
remarkably,
finds signatures of a BOW phase beyond 
the weak-coupling regime where the method is 
under control.

An intermediate BOW phase can also be found in the ionic Hubbard model
\cite{PhysRevLett.83.2014, FABRIZIO2000647, PhysRevB.63.235108, PhysRevB.64.121105, 2003JPCM...15.5895K, PhysRevB.67.205109, PhysRevLett.92.246405, PhysRevB.70.155115, PhysRevB.71.155105, PhysRevB.79.165109} which is subject to a 
staggered chemical potential $\Delta \, (-1)^i \fden{i}$. However, its BOW--CDW transition
is in a different universality class, as the CDW phase does not break a symmetry of the model.
A staggered potential $\Q{i} \to \Delta \, (-1)^i / g$ also describes the ground state of the Hubbard-Holstein model in the adiabatic phonon limit, \ie, $\omega_0 \to 0$ at fixed $k$, for which the CDW order parameter $\Delta$ has to be determined self-consistently in the presence of the local phonon potential $\Delta^2/(2\lambda W)$. At strong coupling, the SDW--CDW transition is first-order, whereas for $\lambda \ll 1$ 
the possibility of an intermediate phase is an open question
\cite{Fehske:2003aa}.

\textit{Conclusions.}---%
We have discovered a narrow intermediate BOW phase in the 1D Hubbard-Holstein model using exact QMC simulations. The BOW--CDW critical point is consistent with a second-order transition described by an LEL fixed point with $K_\rho < 1$, which is likely to emerge from an extended weak-coupling metallic phase and only gives way to a first-order transition at strong couplings. Our mechanism for BOW order goes beyond the weak-coupling Peierls argument
and represents a retardation-driven analog of the extended Hubbard model, leading to  
a strong-coupling phase diagram that is surprisingly similar to this purely electronic model; this requires
us to reconsider the role of phonons in stabilizing exotic phases of matter.
To discover the BOW phase,
it was necessary to change our perspective towards strong couplings $U\approx\lambda W$, but our results should also motivate us
to reconsider the weak-coupling metallic regime, 
its phase boundaries, and how the BOW phase emerges from it. 
It also remains open how the phonon frequency affects the phase structure and the scaling in the critical regime.

\begin{acknowledgments}
\textit{Acknowledgments.}---%
I am very grateful to Fakher Assaad and  Martin Hohenadler for many insightful discussions on the Hubbard-Holstein model during my PhD studies several years ago.
I also acknowledge discussions with Sebastian Paeckel.
This work was supported by the Deutsche Forschungsgemeinschaft
through the W\"urzburg-Dresden Cluster of Excellence on Complexity and Topology
in Quantum Matter---\textit{ct.qmat} (EXC 2147, Project No. 390858490).
The authors gratefully acknowledge the computing time made available to them on the high-performance computer at the NHR Center of TU Dresden. This center is jointly supported by the Federal Ministry of Education and Research and the state governments participating in the NHR \footnote{\url{https://www.nhr-verein.de/unsere-partner}}.
\end{acknowledgments}

\clearpage

\stepcounter{myequation}
\stepcounter{myfigure}

\setcounter{secnumdepth}{3}  

\renewcommand{\thefigure}{S\arabic{figure}}
\renewcommand{\thesection}{S\arabic{section}}
\renewcommand{\thetable}{S\arabic{table}}
\renewcommand{\theequation}{S\arabic{equation}}

\onecolumngrid

\centerline{\bf\large Supplemental Material} \vskip3mm
\centerline{\bf\large for} \vskip3mm
\centerline{\bf\large Intermediate bond-order-wave phase and nature of the order-to-order transition} \vskip1.0mm
\centerline{\bf\large in the one-dimensional Hubbard-Holstein model} \vskip1cm

\maketitle

\section{Signatures of the SDW--BOW--CDW transitions in different observables}
\label{Sec:Stiffness}

A common observable to distinguish gapless from gapped phases in 1D systems is the spin/charge stiffness. Consider a ring of length $L$ threaded by a magnetic flux $\phi_\sigma$ for each spin sector which can be incorporated via twisted boundary conditions $\fan{L+1, \sigma} = e^{\im \phi_\sigma} \fan{1,\sigma}$ in the Hamiltonian. If we define $\phi_\mathrm{s/c} = \phi_\uparrow = \mp \phi_\downarrow$, the spin/charge stiffness
\begin{align}
\rho_\mathrm{s/c}
=
\left. L \frac{\partial^2 F(\phi_\mathrm{s/c})}{\partial \phi_\mathrm{s/c}^2} \right|_{\phi_\mathrm{s/c}=0}
=
\frac{L}{\beta} \left[ \langle{W_\mathrm{s/c}^2}\rangle - \langle{W_\mathrm{s/c}}\rangle^2 \right]
\end{align}
is given by the second derivative of the free energy $F = -\frac{1}{\beta} \ln Z$ with respect to the flux $\phi_\mathrm{s/c}$. In our QMC simulations, $\rho_\mathrm{s/c}$ can be accessed from the fluctuations of the winding number $W_\mathrm{s/c}= W_\uparrow \mp W_\downarrow$. At zero temperature, we have $\rho_\mathrm{s/c} >0$ if the spin/charge sector is gapless and $\rho_\mathrm{s/c} =0$
if it is gapped. $\rho_\mathrm{s/c}$ measures if an infinitesimal flux induces a spin/charge current; therefore $\rho_\mathrm{c}$ corresponds to the Drude weight at zero temperature.

The presence of a spin/charge gap can also be detected from the fluctuations of the total spin/charge operator.
In our QMC simulations, the best estimates are obtained from the long-wavelength limit of the corresponding susceptibilities,
\begin{align}
\kappa_\mathrm{s/c}(L) = \chi_{\sigma/\rho}(q_1) \, ,
\qquad q_1=2\pi/L \, ,
\end{align}
for which $\kappa_\mathrm{s/c} = \lim_{L\to\infty} \kappa_\mathrm{s/c}(L)$ gives the spin/charge compressibility. The spin/charge compressibility has similar properties to the stiffness: it scales to zero if the spin/charge channel is gapped and remains finite if it is gapless.

\begin{figure}[b]
    \centering
    \includegraphics[width=\linewidth]{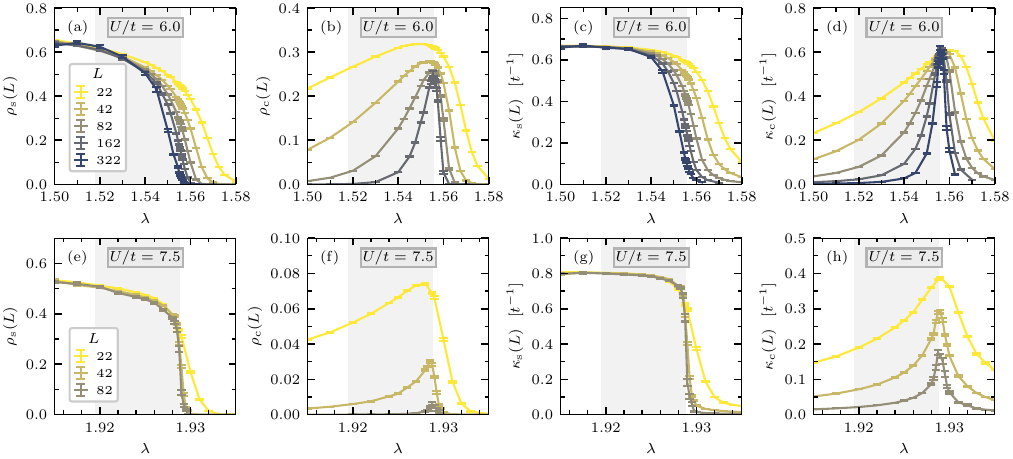}
    \caption{Finite-size analysis of the (a) spin stiffness $\rho_\mathrm{s}(L)$, (b) charge stiffness $\rho_\mathrm{c}(L)$, (c) spin compressibility $\kappa_\mathrm{s}(L)$, and (d) charge compressibility $\kappa_\mathrm{c}(L)$ for $U/t = 6.0$.
  The shaded areas indicate the intermediate BOW phase, as determined in the main paper.
  (e)--(h) Same observables for $U/t = 7.5$.
    }
    \label{figsm:stiffness_compressibility}
\end{figure}

Figure \ref{figsm:stiffness_compressibility} shows the spin/charge stiffness as well as the spin/charge compressibility for $U/t=6.0$ and $U/t = 7.5$.
$\rho_\mathrm{c}(L)$ and $\kappa_\mathrm{c}(L)$ develop a sharp peak at the BOW--CDW phase boundary,
as it is also the case for $K_\rho(L)$ shown in the main paper. The maximum of the peak scales to a finite value for $U/t=6.0$, indicating gapless charge excitations, and to zero for $U/t = 7.5$ as expected for a finite charge gap. At the same time, $\rho_\mathrm{s}(L)$ and $\kappa_\mathrm{s}(L)$ scale to zero at $\lambda = \lambda_{c_2}$ indicating a finite spin gap. For $U/t =6.0$, the presence of a spin gap can be inferred from $\rho_\mathrm{s}(L)$ and $\kappa_\mathrm{s}(L)$ even for $\lambda \lesssim \lambda_{c_2}$. 
However, $\rho_\mathrm{s}(L)$ and $\kappa_\mathrm{s}(L)$ do not give a reliable estimate of $\lambda_{c_1}$ where these quantities are supposed to develop a jump. 
Because the SDW--BOW transition is expected to be a BKT transition, 
the spin gap is exponentially small for $\lambda \gtrsim \lambda_{c_1}$ making it impossible to resolve the spin gap using the system sizes available to our QMC study. As the intermediate BOW regime gets more narrow for $U/t=7.5$, it is hard to see any drop of $\rho_\mathrm{s}(L)$ and $\kappa_\mathrm{s}(L)$ until the system exhibits the first-order BOW--CDW transition at $\lambda = \lambda_{c_2}$. The same difficulties have been observed for the extended Hubbard model \cite{PhysRevB.65.155113}. The drop of the Luttinger parameter $K_\sigma(L)$ below one, as used in the main paper, gives a better estimate of the SDW--BOW phase boundary \cite{PhysRevB.65.155113}, because spin rotational symmetry requires $K_\sigma = 1$ in the SDW phase and $K_\sigma = 0$ in the spin-gapped phases.

\section{Estimation of the Luttinger parameters}
\label{Sec:LuttParam}

In Luttinger-liquid theory, the Luttinger parameters are commonly estimated from the static structure factors via
$K_{\rho/\sigma} = \lim_{q\to 0} \pi S_{\rho/\sigma}(q)/q$. This relation can be derived from the Fourier transform of the real-space correlations, for which the $q=0$ contribution reads $S_{\rho/\sigma, q=0}(r) = - K_{\rho/\sigma}/(\pi r)^2$, but it can also be derived from relations between the corresponding compressibilities and velocities; for details see Ref.~\cite{PhysRevB.59.4665}.
\begin{figure}
    \centering
    \includegraphics[width=\linewidth]{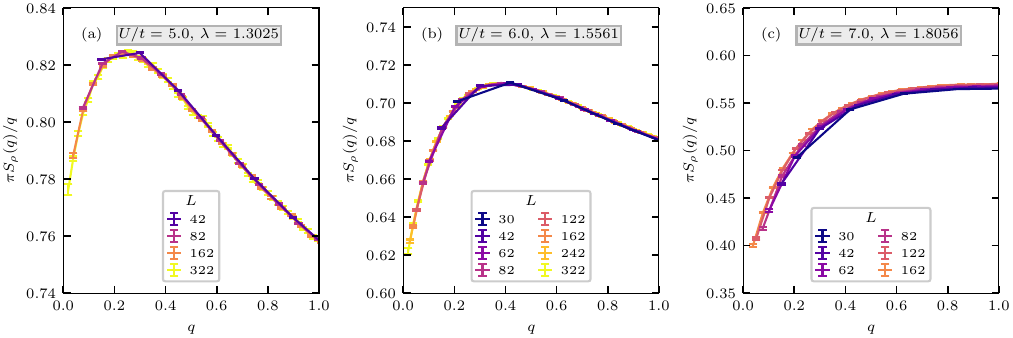}
    \caption{%
Finite-size dependence of $\pi S_\rho(q) / q$ at the BOW--CDW phase boundary $\lambda \approx \lambda_{c_2}$ for $U/t\in\{ 5.0, 6.0, 7.0 \}$.
    }
    \label{figsm:KrhoFS}
\end{figure}
In numerical simulations, it is convenient to estimate $K_{\rho/\sigma}(L) = \pi  S_{\rho/\sigma}(q_1)/q_1$ for a given system size $L$ at the smallest nonzero momentum $q_1=2\pi/L$ and then scale $L\to\infty$.
To check that both definitions are consistent with each other, we show $\pi S_\rho(q)/q$ at the BOW--CDW phase boundary for different system sizes and different Hubbard interactions in Fig.~\ref{figsm:KrhoFS}.
Our results for $U/t=5.0$ and $U/t=6.0$ collapse on top of each other for different $L$, such that the extrapolation at small $q$ is equivalent to $K_\rho(L)$. Only at $U/t=7.0$, the different curves show a small drift with system size over a wide range of $q$ values. Nonetheless, $K_\rho(L)$ still faithfully represents the evolution of $\pi S_\rho(q)/q$ for $q\to 0$. Because $U/t=7.0$ is expected to lie close to a multicritical point where the BOW--CDW transition changes from second- to first-order,
finite-size corrections might be enhanced, as reported for the extended Hubbard model \cite{PhysRevB.65.155113, PhysRevLett.92.236401}.

At the LEL fixed point, a precise extrapolation of $K_\rho(L\to\infty)$ is challenging because the form of the fitting function is not known. This is true even for the attractive Hubbard model, \ie, the $\omega_0 \to \infty$ limit of the Holstein model, where a Luther-Emery liquid with $K_\rho=1$ is known to exist from the exact Bethe-ansatz solution. However, for any finite $L$ that is accessible to QMC simulations, $K_\rho(L)$ is slightly larger than one and only scales towards one very slowly, such that a naive extrapolation would overestimate $K_\rho$ by a few percent \cite{PhysRevB.92.245132}. For our simulations at a finite phonon frequency, the extrapolation of $K_\rho(L)$ remains challenging. In particular, system sizes need to be large enough to cross the peak in $\pi S_\rho(q)/q$ and then we need to extrapolate $q\to0$ from a rather steep decline. For the LEL phase, it has been reported that finite-size corrections of observables in the charge sector are expected to be strong when coupled to phonons \cite{PhysRevB.84.165123}.

\section{Real-space correlation functions}
\label{Sec:Realspace}

\begin{figure}[t]
    \centering
    \includegraphics[width=\linewidth]{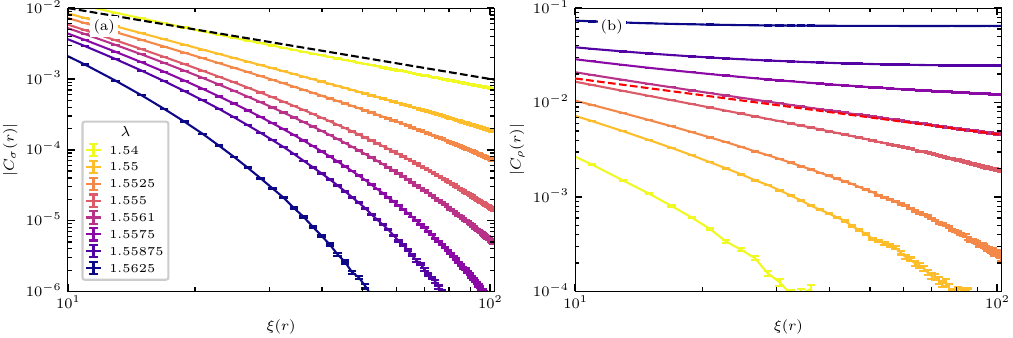}
    \caption{%
Real-space correlation functions $C_{\sigma / \rho} (r)$ in the (a) spin and (b) charge sector for $L=322$ and different $\lambda$
as a function of the conformal distance $\xi(r) = L \sin (\pi r / L) /\pi$. The black dashed line in (a) corresponds to $r^{-1}$
and the red dashed line in (b) to $ r^{-K_{\rho}}$ with $K_\rho = 0.59(1)$.
Here, $U/t = 6.0$.
    }
    \label{figsm:realspace}
\end{figure}

Figure \ref{figsm:realspace} shows the real-space correlation functions $C_{\sigma/\rho}(r)$ for $U/t = 6.0$ and $L=322$. Here, we use the conformal
distance $\xi(r) = L \sin (\pi r / L) /\pi$ to get rid of boundary effects. $C_{\sigma/\rho}(r)$ exhibits the same features as $C_{\sigma/\rho}(r=L/2)$ discussed in the main paper: $C_\sigma(r)$ decays exponentially for $\lambda > \lambda_{c_1}$ but slowly approaches the $1/r$ decay of the SDW phase as $\lambda \to \lambda_{c_1}$. By contrast, $\left| C_\rho(r) \right|$ approaches a constant for $\lambda > \lambda_{c_2}$ and decays exponentially for $\lambda < \lambda_{c_2}$. For the LEL fixed point at the BOW--CDW phase boundary, we expect $C_\rho(r) \propto (-1)^r r^{-K_\rho}$ using the prediction of bosonization for the $q=2\kF$ part of the real-space correlations. At $\lambda = 1.5561 \approx \lambda_{c_2}$, our data is consistent with a power-law decay, although there seems to be a slow drift of the exponent.

\begin{figure}
    \centering
    \includegraphics[width=\linewidth]{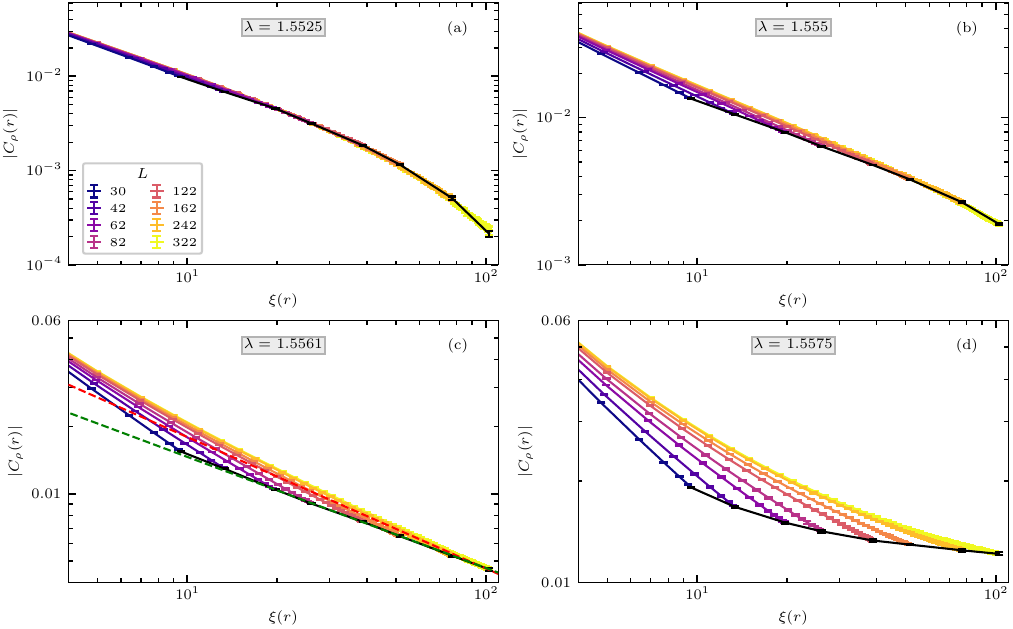}
    \caption{%
Finite-size analysis of the real-space charge correlation function $\left| C_\rho(r) \right|$ for different electron-phonon couplings $\lambda$ across the BOW--CDW transition.
We plot $\left| C_\rho(r) \right|$ as a function of the conformal distance $\xi(r) = L \sin (\pi r / L) /\pi$ to remove boundary effects.
The black data points correspond to $\left| C_\rho(r=L/2) \right|$ shown in the main paper. The dashed lines in (c) illustrate an $r^{-K_\rho}$ decay with $K_\rho=0.59(1)$ (red), as extrapolated from $K_\rho(L)$ in Fig.~2(b), and $K_\rho = 0.50(1)$ (green),
fitting the long-distance decay of $C_\rho(r=L/2)$.
Here, $U/t = 6.0$.
    }
    \label{figsm:realspace_FS}
\end{figure}

To get a better idea on how finite-size effects influence the correlation functions, we plot $\left| C_\rho(r) \right|$ for different system sizes $L$ in Fig.~\ref{figsm:realspace_FS} and compare it to the size dependence of $\left| C_\rho(r=L/2) \right|$. Figure~\ref{figsm:realspace_FS}(a) shows  $\left| C_\rho(r) \right|$ in the BOW phase where we expect to find an exponential decay. At small distances, $\left| C_\rho(r) \right|$ is consistent with a power law, but eventually crosses over towards an exponential decay for longer distances.
Then, our results for different system sizes fall on top of each other and $\left| C_\rho(r=L/2) \right|$ captures the decay of $\left| C_\rho(r) \right|$ faithfully. If we tune $\lambda$ closer to the BOW--CDW transition but still remain in the BOW phase, as shown in Fig.~\ref{figsm:realspace_FS}(b), the short-distance power law extends to larger $r$ with an $L$-dependent exponent that is slightly reduced before $\left| C_\rho(r) \right|$ crosses over towards a faster decay at the longest distances. At $\lambda \approx \lambda_{c_2}$,
this crossover behavior makes it difficult to get a reliable estimate of the asymptotic power-law exponent. Figure \ref{figsm:realspace_FS}(c) shows that for our largest system size $\left| C_\rho(r) \right|$ gets close to a power-law decay with $K_\rho = 0.59(1)$ estimated from $K_\rho(L\to\infty)$, whereas $\left| C_\rho(r=L/2) \right|$ appears to decay more slowly. Figure~\ref{figsm:realspace_FS}(c) suggests that significantly larger system sizes are needed to get a reliable estimate of $K_\rho$ from the correlation functions. Similar finite-size effects are expected for the imaginary-time correlation functions and therefore also for the susceptibilities.
Eventually, Fig.~\ref{figsm:realspace_FS}(d) shows $\left| C_\rho(r) \right|$ in the CDW phase, but not too far away from $\lambda_{c_2}$. Here, the small but finite order parameter can only be resolved on sufficiently large systems, hence we see a strong drift of $\left| C_\rho(r) \right|$ with increasing $L$ and $r$. 
The drift towards a constant order parameter is well captured by $\left| C_\rho(r=L/2) \right|$.

\section{Results for additional Hubbard interactions}
\label{Sec:Add}

\begin{figure}
    \centering
    \includegraphics[width=\linewidth]{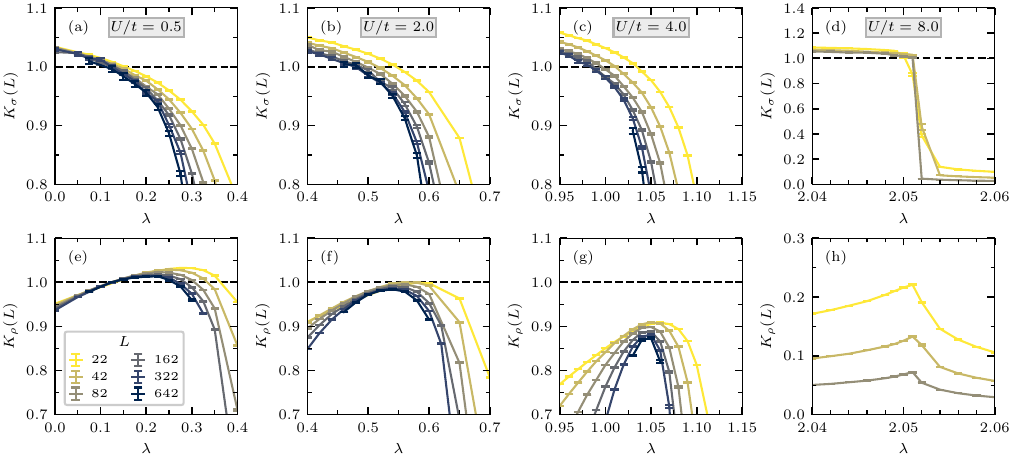}
    \caption{%
    Finite-size analysis of the (a)--(d) spin and (e)--(h) charge Luttinger parameters $K_{\sigma/\rho}(L)$ for Hubbard repulsions $U/t \in \{ 0.5, 2.0, 4.0, 8.0 \}$, as noted on top of each column.
    }
    \label{figsm:Luttinger}
\end{figure}

In our main paper, we have shown that an intermediate BOW phase exists for $U/t \in\{5.0, 6.0, 7.0, 7.5 \}$.
In the following, we present additional results for $K_{\sigma/\rho}(L)$, as displayed in Fig.~\ref{figsm:Luttinger} for $U/t \in\{0.5, 2.0, 4.0, 8.0 \}$.
Figures~\ref{figsm:Luttinger}(d) and \ref{figsm:Luttinger}(h) suggest that the intermediate phase is absent for $U/t=8.0$ (and beyond) because the discontinuity in $K_\sigma(L)$ and the peak in $K_\rho(L)$ appear at the same $\lambda_{c} = 2.051(1)$. The resulting SDW--CDW transition is strongly first-order. However, we expect the BOW phase to exist even for $U/t < 5.0$. At $U/t=4.0$, $K_\rho(L)$ clearly develops a peak that gets more narrow with $L$, as can be seen in Fig.~\ref{figsm:Luttinger}(g), and appears at a distinct position from where $K_\sigma(L)$ drops below one in Fig.~\ref{figsm:Luttinger}(c).
As the Hubbard repulsion gets weaker, the peak in $K_\rho(L)$ covers a significantly broader range of electron-phonon couplings $\lambda$ and the scaling with system size becomes even slower. At $U/t =2.0$ in Fig.~\ref{figsm:Luttinger}(f), we still find $K_\rho(L)<1$ for all $\lambda$, which would be a signature of an isolated LEL point with $K_\rho < 1$ based on the field-theory understanding of the extended Hubbard model.
At $U/t =0.5$ in Fig.~\ref{figsm:Luttinger}(e), we find a regime with $K_\rho(L)>1$, as in the attractive Hubbard model where $K_\rho(L)$ scales to one (from above and extremely slowly) within an extended LEL phase.
Because previous studies for small to intermediate system sizes reported inconsistent behavior for $K_\rho(L)$ and the decay of the correlation functions in the LEL phase,
it has been questioned whether the Luttinger relations can get renormalized by a coupling to phonons \cite{PhysRevB.75.245103, PhysRevB.92.245132} (whereas spin correlations are not expected to be affected because the phonons only couple to the charge sector). The weak-coupling regime is difficult to analyze numerically, because the expected crossover from the Luttinger-liquid to the Luther-Emery-liquid fixed point is very slow.
We will not be able to resolve this issue here, but refer to Refs.~\cite{PhysRevB.75.245103, PhysRevB.92.245132} for a detailed discussion of previous results.

\end{document}